\documentclass[aip,jap,reprint,superscriptaddress,groupeaddress]{revtex4-1}
\usepackage{graphicx}
\usepackage{hyperref}
\usepackage{amssymb}
\begin{document}
\title{Charge transport and electron-hole asymmetry in low-mobility graphene/hexagonal boron nitride heterostructures}

\author{Jiayu Li}
\affiliation{Beijing Key Laboratory of Quantum Devices, Key Laboratory for the Physics and Chemistry of Nanodevices and Department of Electronics, Peking University, Beijing 100871, China}
\affiliation{Academy for Advanced Interdisciplinary Studies, Peking University, Beijing 100871, China}
\author{Li Lin}
\affiliation{Center for Nanochemistry, Beijing Science and Engineering Center for Nanocarbons, Beijing National Laboratory for Molecular Sciences, College of Chemistry and Molecular Engineering, Peking University, Beijing 100871, P. R. China}
\author{Guang-Yao Huang}
\affiliation{Beijing Key Laboratory of Quantum Devices, Key Laboratory for the Physics and Chemistry of Nanodevices and Department of Electronics, Peking University, Beijing 100871, China}
\author{N. Kang}
\email[Corresponding author. ]{nkang@pku.edu.cn}
\affiliation{Beijing Key Laboratory of Quantum Devices, Key Laboratory for the Physics and Chemistry of Nanodevices and Department of Electronics, Peking University, Beijing 100871, China}
\author{Jincan Zhang}
\affiliation{Center for Nanochemistry, Beijing Science and Engineering Center for Nanocarbons, Beijing National Laboratory for Molecular Sciences, College of Chemistry and Molecular Engineering, Peking University, Beijing 100871, P. R. China}
\author{Hailin Peng}
\affiliation{Center for Nanochemistry, Beijing Science and Engineering Center for Nanocarbons, Beijing National Laboratory for Molecular Sciences, College of Chemistry and Molecular Engineering, Peking University, Beijing 100871, P. R. China}
\author{Zhongfan Liu}
\affiliation{Center for Nanochemistry, Beijing Science and Engineering Center for Nanocarbons, Beijing National Laboratory for Molecular Sciences, College of Chemistry and Molecular Engineering, Peking University, Beijing 100871, P. R. China}
\author{H.~Q.~Xu}
\email[Corresponding author. ]{hqxu@pku.edu.cn}
\affiliation{Beijing Key Laboratory of Quantum Devices, Key Laboratory for the Physics and Chemistry of Nanodevices and Department of Electronics, Peking University, Beijing 100871, China}
\affiliation{Division of Solid State Physics, Lund University, P. O. Box 118, S-221 00 Lund, Sweden}

\date{\today}

\begin{abstract}
Graphene/hexagonal boron nitride (G/$h$-BN) heterostructures offer an excellent platform for developing nanoelectronic devices and for exploring correlated states in graphene under modulation by a periodic superlattice potential. Here, we report on transport measurements of nearly $0^{\circ}$-twisted G/$h$-BN heterostructures. The heterostructures investigated are prepared by dry transfer and thermally annealing processes and are in the low mobility regime (approximately $3000~\mathrm{cm}^{2}\mathrm{V}^{-1}\mathrm{s}^{-1}$ at 1.9 K). The replica Dirac spectra and Hofstadter butterfly spectra are observed on the hole transport side, but not on the electron transport side, of the heterostructures. We associate the observed electron-hole asymmetry to the presences of a large difference between the opened gaps in the conduction and valence bands and a strong enhancement in the interband contribution to the conductivity on the electron transport side in the low-mobility G/$h$-BN heterostructures. We also show that the gaps opened at the central Dirac point and the hole-branch secondary Dirac point are large, suggesting the presence of strong graphene-substrate interaction and electron-electron interaction in our G/$h$-BN heterostructures. Our results provide additional helpful insight into the transport mechanism in G/$h$-BN heterostructures.

\end{abstract}

\maketitle
\section{Introduction}
Recent developments in preparing graphene/hexagonal boron nitride (G/$h$-BN) heterostructures by transfer \cite{natnano5.722} and/or growth methods \cite{natmater12.792} have attracted a great deal of attention. \cite{nature497.598,nature497.594,science340.1427,science350.1231,nanolett16.2387} $h$-BN has been used as a substrate and/or an encapsulation layer for graphene in order to minimize the detrimental influence of conventional substrates on the transport properties of graphene and to provide a clean interface in a graphene heterostructure. \cite{natmater10.282} It has been demonstrated that due to a significant reduction in scattering from charged surface states, surface roughness, and impurities of the substrates, many-body correlated states can be observed in high-quality G/$h$-BN heterostructure samples. \cite{prl111.266801,natphys2017,arXiv1611.07113}  It has also been shown that the difference in lattice constant between graphene and $h$-BN leads to the formation of a hexagonal moir\'e pattern, \cite{natmater10.282,natphys8.382,natphys10.451} with a period determined distinctly by the rotation angle between the graphene and $h$-BN lattices. The moir\'e pattern introduces a periodic superlattice potential to the charge carriers in graphene and thus gives rise to a modification of the electronic spectrum and the cloning of the Dirac point. \cite{nature497.598,nature497.594,science340.1427,science350.1231,nanolett16.2387} When the relative rotation angle between the  graphene and $h$-BN lattices is close to zero degree, the moir\'e pattern period can reach $\sim 15~\mathrm{nm}$, enabling the access to the fractal energy spectrum, namely the Hofstadter butterfly spectrum. \cite{prb14.2239}

Although much work has previously been reported on the investigation of the transport properties of G/$h$-BN heterostructures, understandings of the mechanism of large gap opening at the Dirac point and the presence of substantial electron-hole asymmetry still remain as challenging tasks. \cite{nature497.598,nature497.594,science340.1427,science350.1231,nanolett16.2387,prb91.245422} Various theoretical explanations for the observed gap formation and electron-hole asymmetry have been proposed, including, e.g., inversion symmetry breaking, \cite{prb90.155406} the strain effect, \cite{prb89.201404,natcommun6.6308} and many-body interaction. \cite{prl111.266801} In particular, it has been suggested that both interband and intraband scattering have the effects on electron-hole asymmetry seen in the charge transport property measurements, with their relative importance strongly depending on the mobility in the graphene sheets. \cite{prb91.245422} So far, transport measurements of G/$h$-BN heterostructures have been limited to the high-mobility samples. Measurements of the charge transport properties in low-mobility rotationally aligned G/$h$-BN samples have rarely been reported.

In this paper, we report on the electrical transport measurements of low-mobility (approximately $3000 ~\mathrm{cm}^{2}\mathrm{V}^{-1}\mathrm{s}^{-1}$ at 1.9 K) G/$h$-BN heterostructures with a nearly ${{0}^{\circ }}$ lattice rotational angle. The G/$h$-BN heterostructures are fabricated by a dry transfer method \cite{acsnano10.2922} and thermal annealing treatment, the latter being required to obtain a nearly ${{0}^{\circ }}$-rotationally aligned heterostructure. \cite{prl116.126101} We observe the cloning of the Dirac point on the hole transport side, but not on the electron transport side. The physical origin of the observed electron-hole asymmetry is analyzed and it is shown that the asymmetry arises from a large difference between the gaps opened in the conduction and valence bands, and a strong enhancement in the interband contribution to the conductivity on the electron transport side in the low-mobility G/$h$-BN heterostructures. The opened gap values at the central Dirac point and replica Dirac points are also extracted and discussed.

\section{Experiment}

\begin{figure*}
\begin{center}
\includegraphics[width=17cm]{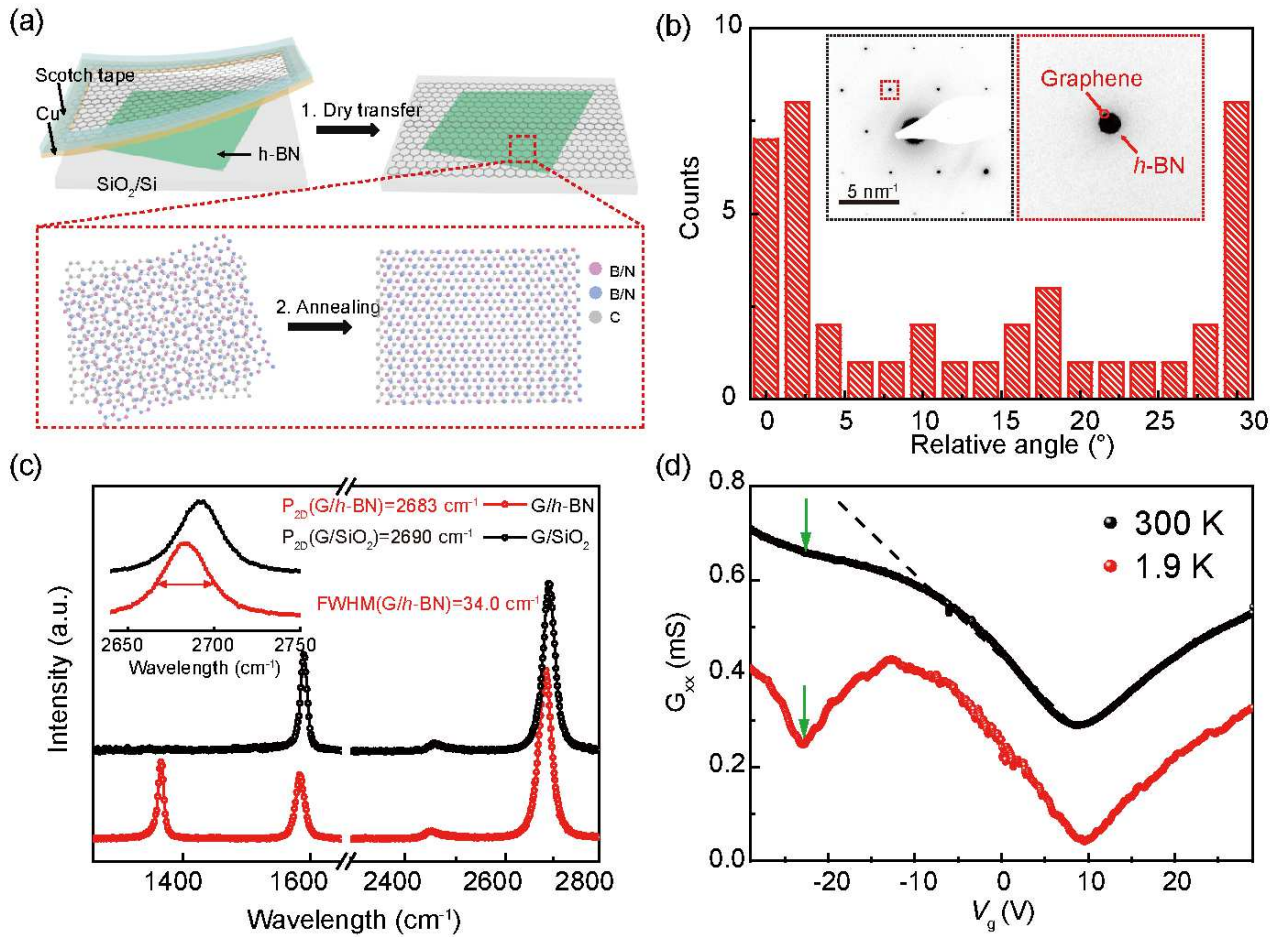}
\caption{(a) Schematic illustration of the fabrication of G/$h$-BN heterostructures with a nearly $0^{\circ}$ twisted lattice angle. (b) Stacking registry survey of G/$h$-BN heterostructures after thermal annealing. (c) Raman spectra of a G/h-BN heterostructure and of a pure graphene sheet on a SiO$_2$ substrate. (d) Conductance versus back gate voltage $V_g$ measured at zero magnetic field at 300 K (black curve) and 1.9 K (red curve).
\label{fig1}}
\end{center}
\end{figure*}

Figure \ref{fig1}(a) schematically illustrates the fabrication of G/$h$-BN heterostructures by the dry transfer method. \cite{acsnano10.2922} First, $h$-BN flakes are mechanically exfoliated onto a Si substrate capped by a 300-nm-thick layer of SiO$_2$ on top with predefined markers for alignments. Subsequently, a 5-$\mathrm{\mu m}$-size monolayer graphene single crystal grown by standard chemical vapor deposition (CVD) is transferred onto an $h$-BN flake using the dry transfer process. \cite{acsnano10.2922,natnanotechnol11.930} This procedure could significantly avoid introducing transfer-related charged impurities and obtain a clear G/$h$-BN interface, enabling a free rotation toward a stable stacking configuration. \cite{prl116.126101} After the transfer, the graphene on $h$-BN is etched into Hall bar geometry using a PMMA mask fabricated by electron beam lithography (EBL) and reactive ion etching (RIE). The sample is then thermally annealed at a temperature of ${{800}^{\circ }}\mathrm{C}$ under pure Ar atmosphere for 12 hours to obtain a nearly ${{0}^{\circ }}$-rotationally aligned heterostructure.\cite{prl116.126101} Selected area electron diffraction (SAED) analysis is performed on a transmission electron microscope (TEM) grid sample with a fabricated G/$h$-BN heterostructure transferred onto it to probe the twist angle between the graphene and $h$-BN lattices. Evidently, as shown in the inset of Fig. \ref{fig1}(b), the SAED pattern obtained for the G/$h$-BN heterostructure after the annealing process exhibits a clear six-fold diffraction structure. A close-up view of the region marked by the red dashed line in the SAED pattern displays two separated spots along a radial direction, where the outer smaller spot reflecting the shorter lattice constant (0.246 nm) and monolayer nature arises from graphene. Figure \ref{fig1}(b) shows a bar graph of the twist angles extracted from the SAED patterns of a collection of G/$h$-BN heterostructures obtained by the dry transfer and thermal annealing process. Here, the dominance of ${{0}^{\circ }}$ and ${{30}^{\circ }}$ twist angles in the stacking registry of the G/$h$-BN heterostructures is seen, which is consistent with the previous work. \cite{prl116.126101} Figure \ref{fig1}(c) shows the measured Raman spectra of a ${{0}^{\circ }}$ lattice rotationally aligned G/$h$-BN heterostructure and pure graphene on the SiO$_2$/Si substrate. According to the experiential rule used in the previous studies of G/$h$-BN heterostructures, \cite{prl116.126101,nanolett13.5242} the full width at half-maximum (FWHM) of the 2D peak can be expressed as FWHM(2D)$=2.5\lambda $, where $\lambda $ is the moir\'e pattern period and thus the 2D peak in a Raman spectrum gives information about the moir\'e pattern period of a G/$h$-BN heterostructure. The FWHM (2D) of the 2D peak in the Raman spectrum of the G/$h$-BN heterostructure shown in figure \ref{fig1}(c) is found to be $\sim 34.0~\mathrm{cm}^{-1}$, corresponding to a moir\'e pattern period of $\sim 14~\mathrm{nm}$. In addition, the 2D band position of the G/$h$-BN heterostructure exhibits a red shift compared to that of the pure graphene sheet on the SiO$_2$/Si substrate, indicating the release of the strain of graphene in the G/$h$-BN heterostructure.

To characterize the electron transport properties of the G/$h$-BN heterostructures obtained by the dry transfer and thermal annealing process, we have fabricated several Hall bar devices (with a width of  $0.8~\mathrm{\mu m}$ and a longitudinal distance between the two probes of  $1.6~\mathrm{\mu m}$). As mentioned above, the Hall bars are defined by EBL and RIE before the thermal annealing. The metal contacts are fabricated, after thermal annealing, by a second step of EBL, evaporation of a metal layer of Ti/Au (5 nm/90 nm in thickness), and lift-off processes. Figure \ref{fig1}(d) depicts the measurements of the longitudinal conductance of a fabricated device as a function of the back gate voltage ${{V}_{g}}$ applied to the Si substrate at zero magnetic field. The transfer curve of the conductance at 300 K (black line) shows a deviation from the linear dependence of the gate voltage (black dashed line) and a noticeable shoulder on the hole side. At 1.9 K (see the red line), another conductance dip appears at ${{V}_{g}}=-23~\mathrm{V}$, i.e., in the region where the conductance shoulder occurs at 300 K, in addition to the main conductance dip at the central Dirac point (CDP, appeared at ${{V}_{g}}=9.8~\mathrm{V}$) of the transfer curve. The point where this additional dip appears is named as a secondary Dirac point (SDP) and is a cloning of the Dirac point in the G/$h$-BN heterostructure. This behavior can be attributed to a large-period moir\'e pattern formed in the heterostructure, which acts as a periodic potential landscape and gives rise to mini-Brillouin-zone bands.\cite{natnano5.722,natmater12.792,nature497.598,nature497.594,science340.1427} The gate voltage difference between the CDP and the SDP in Fig. \ref{fig1}(d) is denoted by $\Delta {{V}_{g}}$, and it is related to the large moir\'e superlattice period ($\lambda$) by the equation
\cite{nature497.598}
\begin{equation}
\Delta {{V}_{g}}=8e/(\sqrt{3}{{\lambda }^{2}}\times{{C}_{g}}),
\end{equation}
where ${{C}_{g}}\sim 10~\mathrm{nF}/\mathrm{c}{{\mathrm{m}}^{2}}$ is the capacitance of the device estimated from the Hall-bar device structure with 5 nm in the thickness of the $h$-BN flakes and 300 nm in the thickness of the SiO$_2$ substrate. Thus, the estimated $\lambda $ is 15.2 nm and the twisting angle is smaller than ${{1}^{\circ }}$, which is consistent with the results of the Raman and TEM characterizations. Compared with the results of previous studies, \cite{nature497.598,nature497.594,science340.1427,science350.1231,nanolett16.2387} in which the SDP in G/$h$-BN heterostructures is observed, the mobility of our sample is rather low ($\sim 2070 ~\mathrm{cm}^{2}\mathrm{V}^{-1}\mathrm{s}^{-1}$ on the electron side and $\sim 2860 ~\mathrm{cm}^{2}\mathrm{V}^{-1}\mathrm{s}^{-1}$ on the hole side at 300 K, whereas $\sim 2980 ~\mathrm{cm}^{2}\mathrm{V}^{-1}\mathrm{s}^{-1}$ on the electron side and $\sim 3940 ~\mathrm{cm}^{2}\mathrm{V}^{-1}\mathrm{s}^{-1}$ on the hole side at 1.9 K). At 300 K, the observation of the SDP in the low-mobility G/$h$-BN sample indicates strong supperlattice potential modulation, which requires both the ${{0}^{\circ }}$-twisted angle between the graphene and $h$-BN lattices and the sufficiently strong interlayer coupling between the graphene sheet and the $h$-BN layer. Here we emphasize again that during the preparation of the G/$h$-BN heterostructures, strong interlayer coupling can be obtained by the thermal annealing process. Before evaporating the 5-nm-Ti/90-nm-Au contact metal layer, we have annealed our samples at a temperature of ${{800}^{\circ }}\mathrm{C}$ under pure Ar atmosphere for 12 hours to remove the remnant contamination and achieve high interface quality.

\section{Results and discussion}

\begin{figure*}
\begin{center}
\includegraphics[width=17cm]{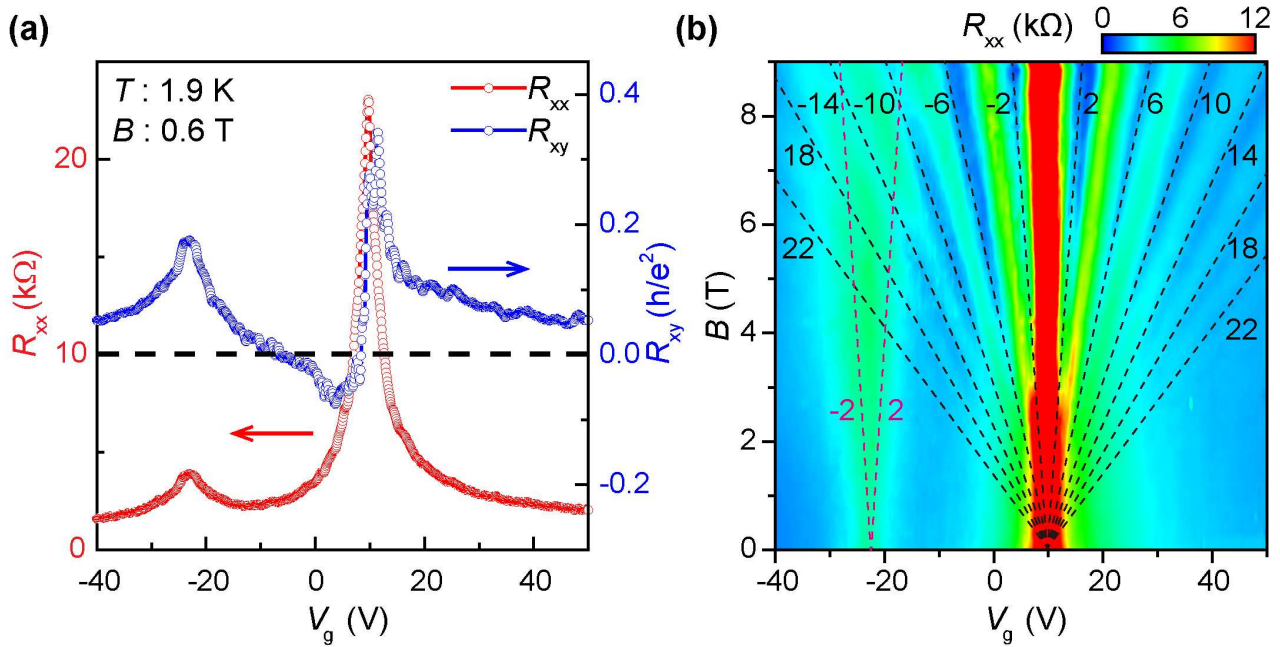}
\caption{(a) Longitudinal resistance $R_{xx}$ (red) and Hall resistance $R_{xy}$ (blue) as functions of back gate voltage $V_g$ at magnetic field $B = 0.6~\mathrm{T}$ and temperature $T = 1.9~\mathrm{K}$. (b) Longitudinal resistance $R_{xx}$ measured as a function of back gate voltage $V_g$ and magnetic field $B$ (Landau fan diagram) at $T = 1.9~\mathrm{K}$.
\label{fig2}}
\end{center}
\end{figure*}

Now, we present and discuss the detailed magnetotransport measurements of the fabricated G/$h$-BN heterostructures. In Fig. \ref{fig2}(a), the measured longitudinal resistance, ${{R}_{xx}}$, and transverse resistance, ${{R}_{xy}}$, are plotted as functions of the gate voltage at magnetic field $B = 0.6~\mathrm{T}$ and temperature $T=1.9$ K. Around the CDP, the Hall resistance changes sign as the Fermi level passes from the valence band to the conduction band. As the Fermi level moves towards the deep inside of the valance band from the CDP, the Hall resistance changes sign again when the gate voltage crosses $-10~\mathrm{V}$. The change in the sign of ${{R}_{xy}}$ around ${{V}_{g}}=-10~\mathrm{V}$ indicates that electron-like carriers appear in the valence band of graphene as a result of the strong modulation of periodic moir\'e potential in the G/$h$-BN heterostructure. Figure \ref{fig2}(b) shows a quantum Hall fan diagram with a color mapping of ${{R}_{xx}}$ as a function of the gate voltage and magnetic field at $T = 1.9~\mathrm{K}$. In Fig. \ref{fig2}(b), we can easily distinguish between the Landau levels (LLs) of the Dirac fermions in terms of the longitudinal resistance minima at filling factors $\nu=4n+2$  (with LL index $n=0,\pm 1,\pm 2,...$) fanning out from the CDP, as indicated by the black dashed lines along the ${{R}_{xx}}$ valleys. \cite{nature438.197,nature438.201} The central ${{R}_{xx}}$ peak fanning out from the CDP corresponds to the zeroth LL, which indicates the gapless Dirac spectrum of monolayer graphene. This distinctly illustrates the unaltered nature of the Dirac fermions in graphene, even under the heavy influence of the moir\'e pattern. The magenta dashed lines in Fig. \ref{fig2}(b) indicates the quantum Hall states associated with the SDP in the hole band. These lines, which fan out from the hole-branch SDP, can be attributed to the quantum Hall states at filling factors $\nu=\pm 2$, whose slope coincides with that of the black dashed lines fanning out from the CDP with $\nu=\pm 2$. The result can be explained by the fact that a half-filled LL at the hole-branch SDP has the same number of carriers as in the half-filled zeroth LL at the CDP. Thus, the LL degeneracy at the SDP equals that at the CDP. This provides a confirmation that the moir\'e superlattice, acting as a periodic potential modulation, gives rise to mini-Brillouin-zone bands.

\begin{figure*}
\begin{center}
\includegraphics[width=17cm]{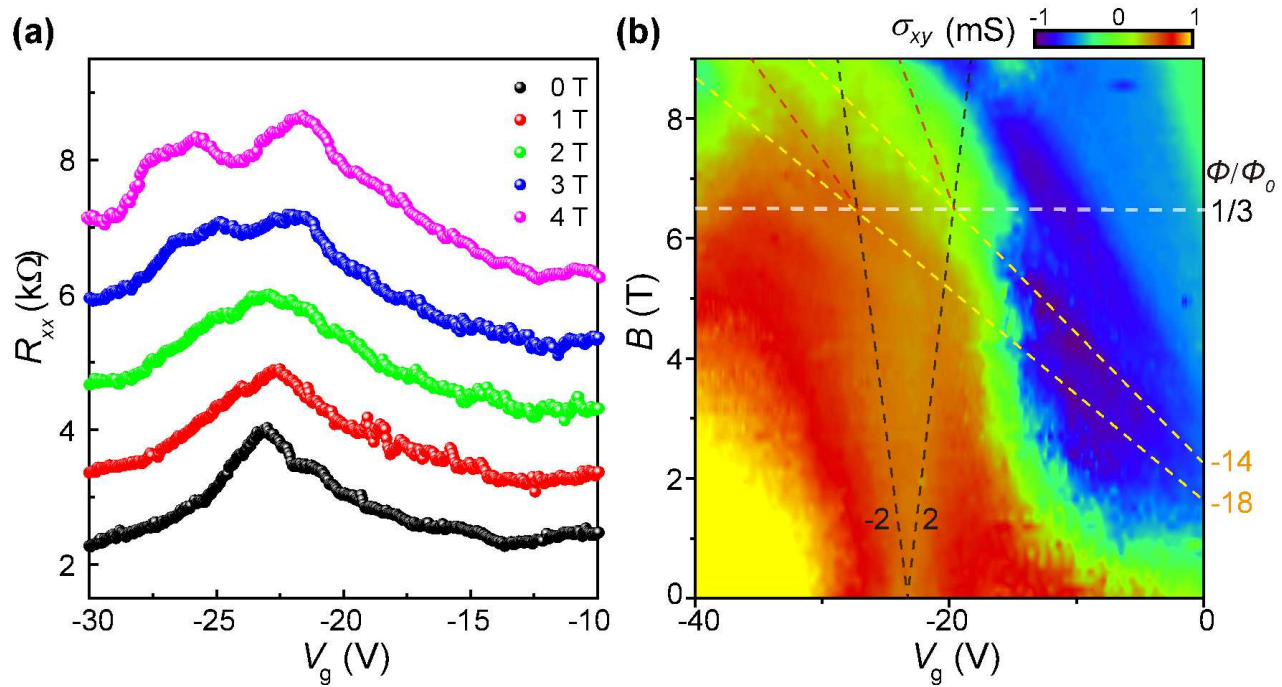}
\caption{(a) Longitudinal resistance $R_{xx}$ as a function of back gate voltage $V_g$ near the hole-branch SDP at different magnetic fields and $T = 1.9~\mathrm{K}$. (b) Hall conductivity $\sigma_{xy}$ as a function of back gate voltage $V_g$ and magnetic field $B$ at $T = 1.9~\mathrm{K}$.
\label{fig3}}
\end{center}
\end{figure*}

Although we have observed the central ${{R}_{xx}}$ peak fanning out from the CDP, we did not observed a central peak of ${{R}_{xx}}$ fanning out from the hole-branch SDP, namely the central ($n=0$) LL, which is vanishing at the hole-side Dirac LL sequence. To demonstrate this more clearly, we plot ${{R}_{xx}}$ as a function of ${{V}_{g}}$ near the SDP under different magnetic fields $B=0$,  1,  2,  3, and 4 T in Fig. \ref{fig3}(a). At the hole-branch SDP, there is only one peak in the ${{R}_{xx}}$ curve at zero magnetic field. The single peak at the hole-branch SDP gradually splits into two peaks as the magnetic field increases. When the magnetic field is set to 4 T, it can be clearly seen that there are two prominent peaks of ${{R}_{xx}}$ on each side of the hole-branch SDP and that the central peak of ${{R}_{xx}}$ at the SDP is vanishing. This result can be understood as that when the graphene and $h$-BN are closely stacked on each other, the graphene $\pi $-electron inherits the inversion symmetry breaking of $h$-BN and the valley degeneracy between $K$ and $K'$ levels in the graphene sheet is lifted, giving rise to splitting into the Landau levels under a finite magnetic field at the SDP.

In Fig.~\ref{fig3}(b), we have plotted a color mapping of the Hall conductivity ${{\sigma }_{xy}}$, calculated from the measured resistances according to ${{\sigma }_{xy}}={{R}_{xy}}/({{\rho }_{xx}}^{2}+{{R}_{xy}}^{2})$. We have labeled the filling factors of the LLs fanned out from the CDP (orange dashed lines) and the hole-branch SDP (black dashed lines) in Fig. \ref{fig3}(b). When a LLs from the CDP meets a LLs from the hole-branch SDP, a new structure appears, which cannot be traced to the LLs from either the CDP or the SDP. For example, at the intersection of the ${\nu_{C\!N\!P}}=-18$ and the ${\nu_{S\!D\!P}}=-2$ LL, a new structure emerges as indicated by a red dashed line. These new structures are in fact the traces of the Hofstadter butterfly spectra as reported in the previous studies. \cite{nature497.598,nature497.594,science340.1427,science350.1231,nanolett16.2387}. In our G/$h$-BN heterostructure, these new structures are seen in Fig.~\ref{fig3}(b) to emerge at B = 6.5 T (where $\phi /{{\phi }_{0}}=1/3$, $\phi $ is the magnetic flux per superlattice unit cell, and ${{\phi }_{0}}=h/e$ is the magnetic flux quantum). The observation of the fractal spectrum at such a low magnetic field suggest a strong moir\'e supperlattice potential modulation in the G/$h$-BN heterostructure. \cite{nanolett16.2387,rmp84.1067}

It is worth noting that the observation of the Hofstadter butterfly spectra and the SDP only at the hole branch in such a low-mobility sample is peculiar. Electron-hole asymmetry in the transport properties of G/$h$-BN heterostructures was usually observed in previous experimental studies. \cite{nature497.598,nature497.594,science340.1427,science350.1231,nanolett16.2387} However, we observe the side peak only on the hole side with no obvious signature of it on the electron side. This is different from the previous reports on high mobility samples in which the side peak appears on both the electron side and the hole side. \cite{nature497.598,nature497.594,science340.1427,science350.1231,nanolett16.2387} The influence of the periodic moir\'e pattern on intersublattice hopping terms in the graphene sheet Hamiltonian gives rise to distinct difference effects on the conduction and valence band near the SDPs. \cite{prb91.245422,prb90.155406} When both the site energy and hopping amplitude distortions are accounted for properly, an overall gap of $\sim 3.5~\mathrm{meV}$ appears in the valence band at the hole-branch SDP, but no such a gap exists in the conduction band. \cite{prb91.245422} By applying the Boltzmann transport theory to the bands predicted by the moir\'e band Hamiltonian, DaSilva et al. have calculated the effects of the interband and intraband responses on the conductivity. \cite{prb91.245422} It has been found that when the charge fluctuation induced Bloch state energy uncertainty, $\hbar /\tau $ (where $\tau $ is the carrier scattering time and is scaled linearly with the carrier mobility), is comparable with the gap at a SDP, the effect of interband scattering is non-negligible. \cite{prb91.245422,natmater10.282} When the mobility reaches $50000 ~\mathrm{cm}^{2}\mathrm{V}^{-1}\mathrm{s}^{-1}$, $\hbar /\tau $ is $\sim 0.1 ~\mathrm{meV}$ and the interband contribution to the conductivity is found to be negligible on both the hole side and the electron side. \cite{prb91.245422,prb82.041406} When the mobility reaches $5000 ~\mathrm{cm}^{2}\mathrm{V}^{-1}\mathrm{s}^{-1}$, $\hbar /\tau $ is around 1 meV, which is smaller than the band splitting of $\sim 3.5 ~\mathrm{meV}$ on the hole side, and therefore the interband scattering is still negligible on the hole side. \cite{prb91.245422} However, the interband contribution at this mobility is found to be non-negligible on the electron side and has a peak at the point where the intraband contribution to the electron conductivity shows a dip feature. In our sample, the mobility is at an even lower value of $\sim 3000 ~\mathrm{cm}^{2}\mathrm{V}^{-1}\mathrm{s}^{-1}$. Thus, on the hole side, the size of the gap is still larger than $\hbar /\tau $ and the interband contribution to the conductivity is still negligible, and the measured conductivity dip at the SDP manifests dominantly the dip in the intraband conductivity. However, on the electron side, the avoided crossing gaps are not typically large compared to $\hbar /\tau $ in our sample, allowing for a peak-like, strong interband contribution at an avoided crossing gap point where the intraband conductivity exhibits a dip feature. Overall, the observed pronounced electron-hole asymmetry in our low-mobility samples can be well explained by the enhanced interband contribution to the conductivity on the electron side.

\begin{figure*}
\begin{center}
\includegraphics[width=17cm]{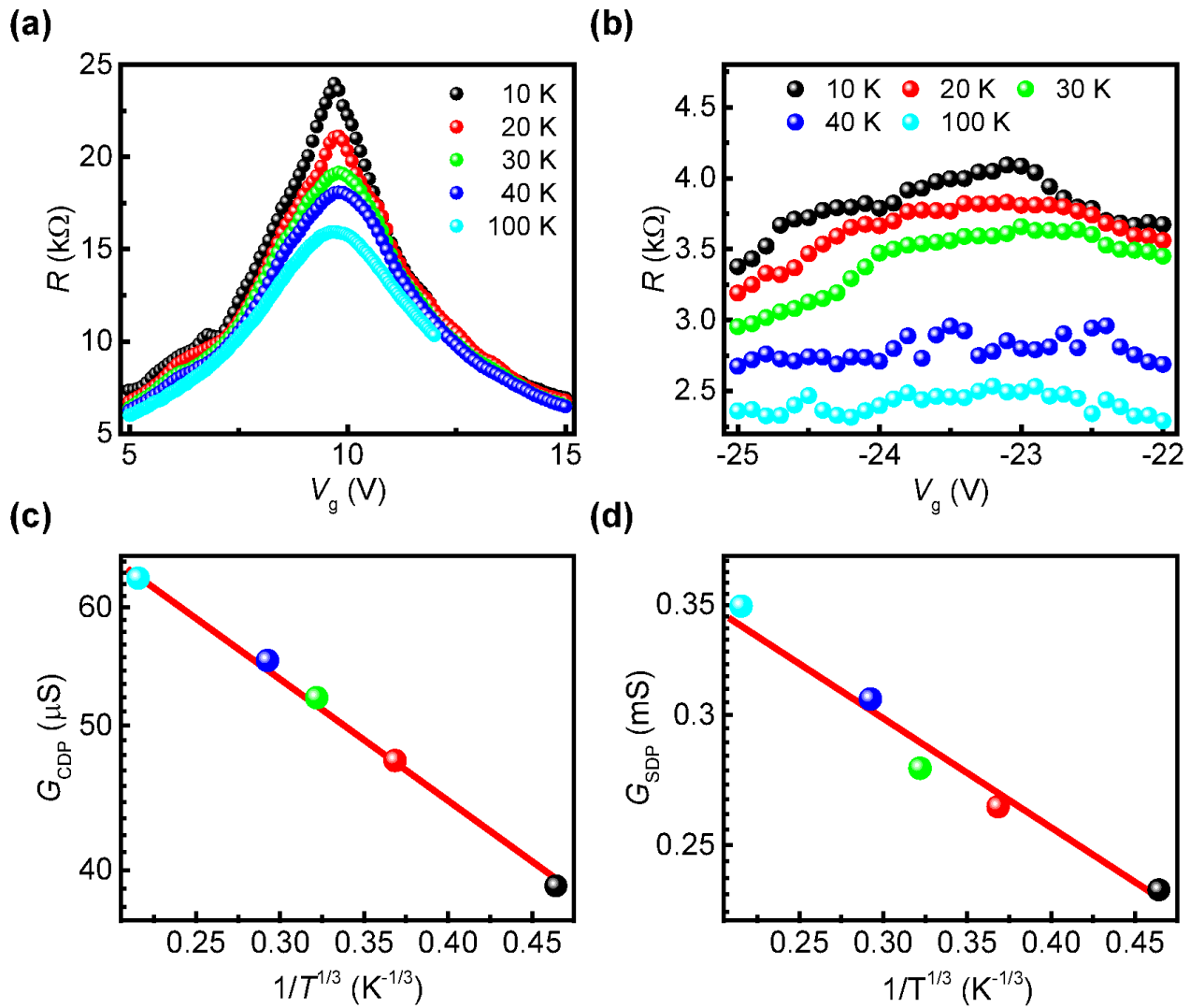}
\caption{(a) Longitudinal resistance $R_{xx}$ as a function of back gate voltage $V_g$ near the CDP at different temperatures. (b) Longitudinal resistance $R_{xx}$ as a function of back gate voltage $V_g$ near the hole-branch SDP at different temperatures. (c) Conductance $G_{xx}$at the CDP plotted on a logarithmic scale versus $T^{-1/3}$. (d) Conductance $G_{xx}$ at the hole-branch SDP plotted on a logarithmic scale versus $T^{-1/3}$.
\label{fig4}}
\end{center}
\end{figure*}

Figure \ref{fig4} shows the resistance measured around both the CDP and the hole-branch SDP at different temperatures. In Figs.~\ref{fig4}(a) and \ref{fig4}(b), the resistance at both the CDP and the hole-branch SDP increases with decreasing temperature, indicative of an insulating behavior. We find that the temperature dependences of the conductance at the CDP and the hole-branch SDP follow the relation of $\ln G\propto -{{T}^{1/3}}$ as seen in Figs.~\ref{fig4}(c) and \ref{fig4}(d), the characteristics of variable-range hopping (VRH) transport in a two-dimensional system. \cite{prb84.045431,prb82.081407} The good agreement of the data with the VRH model indicates that charge transport is dominated by carrier hopping via localized states in our low-mobility samples with potential disorder. \cite{prb82.081407,prl105.166601} The extracted gap at the CDP is $\sim 69 ~\mathrm{K}$, while the extracted gap at the hole-branch SDP is 35 K, which is smaller than the gap at the CDP. A larger energy gap opening at the CDP in our samples is similar to the previous work on high mobility G/$h$-BN heterostructures \cite{nature497.598,nature497.594,science340.1427,science350.1231,nanolett16.2387} and has been attributed to gap enhancements induced by the interaction with the substrate and electron-electron interaction at the CDP. \cite{prl111.266801,prb89.201404,natcommun6.6308} In addition, the annealing process induced well-lattice rotationally aligned G/$h$-BN heterostructures could lead to the formation of the commensurate states in the graphene sheets, offering an alternative explanation for the gap opening at the CDP. \cite{natphys10.451}

\section{Conclusion}
In conclusion, low-mobility rotationally aligned G/$h$-BN heterostructures have been studied by charge transport measurements at low temperatures. The G/$h$-BN heterostructures are fabricated by the dry transfer and thermal annealing techniques. Strong electron-hole asymmetry in the transport properties has been observed in the low-mobility rotationally aligned G/$h$-BN heterostructures. It is shown that the replica Dirac spectra and the signatures of the Hofstadter butterfly spectra are clearly observable on the hole transport side in our low-mobility G/$h$-BN heterostructures. Our experimental results are in good agreement with the recent theoretical calculations and the strong electron-hole asymmetry can be well explained  by the presence of a large difference in gap opening at the SDPs in the conduction and valence bands of the graphene sheet and a strong enhancement in the interband contribution to the conductivity on the electron transport side of the graphene sheet. The gap openings of 69 K and 35 K at the CDP and the SDP in the valence band are also extracted from the temperature-dependent measurements of the conductance. These large gap openings suggest the presence of strong graphene-substrate interaction and electron-electron interaction in our G/$h$-BN heterostructures.

\section*{Acknowledgments}
This work was financially supported by the Ministry of Science and Technology of China through the National Key Research and Development Program of China (Grant Nos.~2016YFA0300601 and 2017YFA0303304), the National Natural Science Foundation of China (Grant Nos.~91221202, 91421303, 11774005 and 11374019), and the Swedish Research Council (VR).

%\clearpage

%\clearpage

%\clearpage

%\clearpage

\end{document}